\documentclass[prd,twocolumn,nofootinbib]{revtex4-2}

\usepackage{graphicx}
\usepackage{color}
\usepackage{amsmath,amssymb}
\usepackage{bbold}
\usepackage{t1enc}

\usepackage{amsmath,amssymb}
\usepackage{amsfonts}
\usepackage{bbold}
\usepackage{latexsym} 
\usepackage{cancel}
\usepackage{graphicx, rotating}
\usepackage{epsfig}
\usepackage{color}
\usepackage[dvipsnames]{xcolor}
\usepackage{multirow}

\DeclareMathSymbol{\shortminus}{\mathbin}{AMSa}{"39}
\DeclareMathSymbol{\shm}{\mathbin}{AMSa}{"39}

\newcommand{\met}{\cancel{E}_T}

\begin{document}

\title{Mass-unspecific classifiers for mass-dependent searches}

\author{J. A.~Aguilar-Saavedra}
\author{S. Rodr\'{\i}guez-Ben\'{\i}tez}
\affiliation{Instituto de F\'isica Te\'orica IFT-UAM/CSIC, c/Nicol\'as Cabrera 13--15, 28049 Madrid, Spain} 

\begin{abstract}
Searches for new particles often span a wide range of mass scales, where the shape of potential signals and the SM background varies significantly. We make use of a multivariate method that fully exploits the correlation between signal and background features and the explored mass scale, and is trained on a sample that is balanced across the entire mass range. The classifiers, either a neural network or a boosted decision tree, produce a continuous output across the full mass range and, at a given mass, achieve nearly the same performance as a classifier specifically trained for that mass. The performance of the classifiers is better than the one obtained with parameterised neural networks and similar methods.
\end{abstract}

\maketitle

\section{Introduction}

The Large Hadron Collider (LHC), its high-luminosity upgrade (HL-LHC) and a potential future circular collider (FCC) will explore increasingly higher scales in the search for new physics. And, at lower scales, feebler interactions will be probed as more statistics are accumulated. A priori it is not known where new particles might manifest: whether at the multi-TeV scale with a coupling of order unity, or at the TeV scale but with a smaller coupling. New physics searches must therefore be sensitive to the presence of new particles in all the allowed parameter space. This cannot be achieved with traditional cut-based methods for a simple reason: while heavier particles produce highly-energetic decay products, lighter particles do not so, and fixed cuts on, say, transverse momenta ($p_T$) of jets and leptons cannot be simultaneously optimised for new particles over a wide mass range.

The ATLAS and CMS experiments have addressed this difficulty by using a parameterised neural network (pNN)~\cite{Baldi:2016fzo}. Following that proposal, the set of features included in a multivariate discriminant is extended with a mass label; for signal samples it corresponds to the mass of the new particle (or new particles, in general), while for the background the value is randomly assigned within the mass interval considered. The pNN classifier is able to learn to interpolate for signal masses not used in the training. 
 In this work we apply the mass unspecific supervised tagging (MUST) concept~\cite{Aguilar-Saavedra:2020uhm} to tackle this issue. MUST was originally developed for efficient tagging of `new physics' jets\footnote{By new physics jets we refer to multi-pronged jets originating from the hadronic decay of a boosted heavy particle.} against the QCD background, over an arbitrary range of jet mass and $p_T$. The key to that goal was the inclusion of jet mass and $p_T$ (both for signal and background) along with subjettiness variables~\cite{Thaler:2010tr,Datta:2017rhs} in a multivariate classifier such as a neural network (NN)~\cite{Aguilar-Saavedra:2020uhm} or boosted decision tree (BDT)~\cite{Aguilar-Saavedra:2023pde}, with subsequent training over all possible jet masses and $p_T$, in samples that are balanced in these two variables. 

For new physics searches using a mass-unspecific classifier, which we dub as $\mu$NN or $\mu$BDT,  the `mass label' to be included in the set of features is the characteristic scale where the new particle is expected to show over the background. This can be the reconstructed particle mass or a transverse mass, for example. 
Such a scale is natural in many experimental analyses that have used a pNN, including
searches for new scalars~\cite{CMS:2019rlz,CMS:2024rgy,ATLAS:2024auw}, vector bosons~\cite{ATLAS:2023vxg,ATLAS:2024uvu}, leptoquarks~\cite{ATLAS:2023uox} and supersymmetric particles~\cite{CMS:2021cox,ATLAS:2022ihe}.
This replacement for the mass label defined in terms of a characteristic mass scale is meaningful for {\em both} the signal and the background. And, most importantly, it captures the differences in the background distributions that arise at different mass scales. Interpolation is not an issue here because, due to detector and reconstruction effects, one has a practically continuous variation of the mass parameter when the signal is generated at discrete points with a separation of the order of the experimental resolution.

Here we take as benchmark process the single production of a vector-like quark singlet $T$ with charge $2/3$~\cite{Aguilar-Saavedra:2013qpa} at the HL-LHC. The cross section is proportional to the square of a mixing angle $|V_{Tb}|^2$, and it also depends on the quark mass $m_T$ via the amplitudes and parton density functions. This process can probe $T$ masses of several TeV with mixing $V_{Tb} \sim 1$, as well as small mixings $V_{Tb}$ for $m_T \sim 1$ TeV, and is suited to test the application of mass-unspecific classifiers for a mass-dependent search.\footnote{Pair production of vector-like quarks, mediated by QCD interactions, has a cross section independent of the heavy quark mixing. It has a smaller mass reach than single production but is able to set constraints on mass alone, independently of the mixing. These constraints are model-dependent and relaxed if additional decay modes are present, for example in models with extra scalars~\cite{Aguilar-Saavedra:2017giu}. Therefore, the exploration of the full range of $T$ masses accessible in single production is well motivated.}
We address signal to background discrimination via a $\mu$NN or a $\mu$BDT, which have the same performance, and compare our results with a pNN and other classifiers. We also show that a $\mu$NN or a $\mu$BDT easily allow to preserve the background shape upon a selection on the classifier output, thereby allowing for a background normalisation from sidebands, if required.

\section{Event generation and pre-selection}

\begin{figure}[htb]
\begin{center}
\includegraphics[height=4cm,clip=]{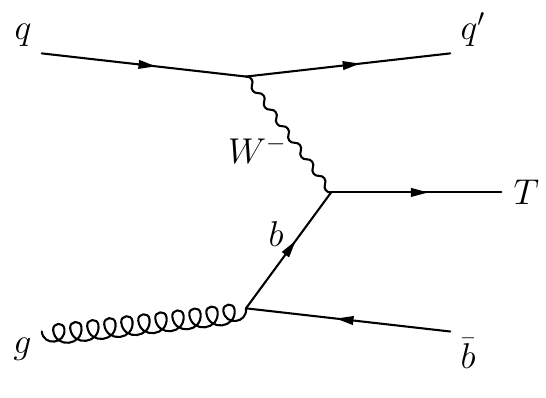} 
\caption{Representative diagram of single $T$ quark production in hadron collisions.}
\label{fig:diag}
\end{center}
\end{figure}

Our signal process is single production of a $T$ quark, $pp \to T \bar b j$, with decay $T \to tZ \to \ell^+ \nu b \ell^+ \ell^-$, $\ell=e,\mu$. Figure~\ref{fig:diag} shows the leading Feynman diagram for the process. As background we include the leading one, which is $p p \to ZW^+ jj$, with $Z \to \ell^+ \ell^-$, $W^+ \to \ell^+ \nu$.\footnote{A more accurate evaluation of the contribution from this background involves matching with $ZW^+$ and $ZW^+j$ processes but this is not relevant to our purposes.} Both processes are generated with {\scshape MadGraph}~\cite{Alwall:2014hca}. For the signal, the $T$ interactions are implemented in {\scshape Feynrules}~\cite{Alloul:2013bka} and interfaced to {\scshape MadGraph} using the universal Feynrules output~\cite{Degrande:2011ua}. Hadronisation and parton showering performed with Pythia~\cite{Sjostrand:2007gs} and detector simulation with Delphes~\cite{deFavereau:2013fsa} using the configuration for HL-LHC. The reconstruction of jets and the analysis of their substructure is done using FastJet~\cite{Cacciari:2011ma}, using the anti-$k_T$ algorithm~\cite{Cacciari:2008gp}. We build two collections of jets, with radius $R=0.8$ and $R=0.4$ respectively.

We restrict ourselves to the kinematical region where the top quark from $T$ decay is boosted, so the $b$ quark and charged lepton are contained within a $R=0.8$ jet.  Signal samples of $1.5 \times 10^5$ events are generated for $T$ masses from 1 TeV to 6.4 TeV in intervals of 200 GeV. The background samples used to train the $\mu$NN and $\mu$BDT, as well as for tests, are generated placing cuts on the invariant mass of the $Z$, $W$ and the jet closest to the $W$ ($m_{ZWj}$). This parton-level quantity serves as a proxy for the reconstructed $T$ mass after hadronisation and detector simulation. Samples of $4 \times 10^5$ events are generated in 200 GeV intervals starting at $0.8 \leq m_{ZWj} \leq 1$ TeV and up to $6.4 \leq m_{ZWj} \leq 6.6$ TeV. With the background generated in this fashion, the entire range of reconstructed $T$ masses up to and above 6 TeV is populated. A second background sample of $6 \times 10^6$ events, generated with a lower cut $m_{ZWj} \geq 800$ GeV, is used for training and validation of the pNN.

Preselected events must have three charged leptons with pseudorapidity $|\eta| \leq 2.5$ and $p_T \geq 25$ GeV, the leading lepton with $p_T \geq 40$ GeV. The $Z$ boson is reconstructed with an opposite-sign same-flavour pair, $p_Z = p_{\ell_1} + p_{\ell_2}$, where we label the most energetic of the two as $\ell_1$. If there are more than one opposite-sign same-flavour pair we select the one with invariant mass closest to the $Z$ boson mass. The reconstructed $Z$ must have $p_T \geq 60$ GeV. The third lepton, labelled as $\ell_3$, is assumed to result from a $W$ boson decay.

We require a $R=0.8$ fat jet $J$ with $p_T \geq 80$ GeV and $|\eta| \leq 2.5$, containing the third lepton $\ell_3$. In the case of the signal this jet corresponds to the top quark. This jet must be $b$-tagged and separated from the reconstructed $Z$ by a distance $\Delta R(J,Z) \geq 1$, where $\Delta R = [ (\Delta \eta)^2 + (\Delta \phi)^2 ]^{1/2}$ has the usual definition.  We also require a $R=0.4$ jet with $p_T \geq 25$ GeV, $|\eta| \leq 5$ that is not $b$-tagged.

The momentum of the escaping neutrino $p_\nu$ is determined assuming it is produced in the decay of a $W$ boson. The transverse components of its momentum are set equal to the missing energy in the event, $(p_\nu)_{x,y} = (\met)_{x,y}$. The kinematical constraint $(p_{\ell_3} + p_\nu)^2 = M_W^2$ provides a second-order equation with two solutions for $(p_\nu)_z$, of which we choose the one that gives a top reconstructed mass $m_t^\mathrm{rec} = [(p_J + p_\nu)^2]^{1/2}$ closest to the top quark mass. If there is no real solution we set the discriminant of the equation to zero.
The reconstructed top momentum is defined as $p_t = p_J + p_\nu$.

The $T$ signal produces a peak in the reconstructed $T$ mass $m_T^\mathrm{rec} = [(p_t + p_Z)^2]^{1/2}$. Additional signal to background discrimination is provided by the following variables:
\begin{itemize}
\item The transverse momenta of the $Z$ boson and $R=0.8$ jet, $p_T^Z$, and $p_T^J$ respectively. 
\item The transverse momenta of the two leptons from $Z$ decay, $p_T^{\ell_1}$, and $p_T^{\ell_2}$.
\item The reconstructed top quark mass.
\item The azimuthal angle between the reconstructed $Z$ and the third lepton, $\Delta \phi(Z,\ell_3)$.
\item The multiplicity of forward $R=0.4$ jets, with $2.5 \leq |\eta| \leq 5$.
\end{itemize}

\begin{figure}[htb]
\begin{center}
\begin{tabular}{c}
\includegraphics[height=6cm,clip=]{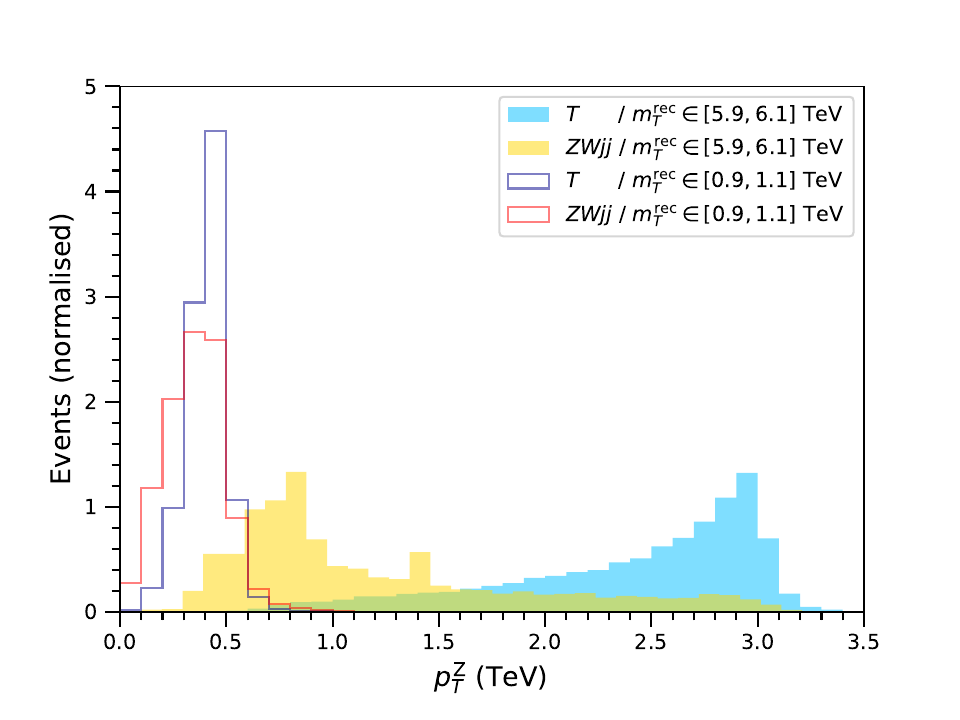} \\
\includegraphics[height=6cm,clip=]{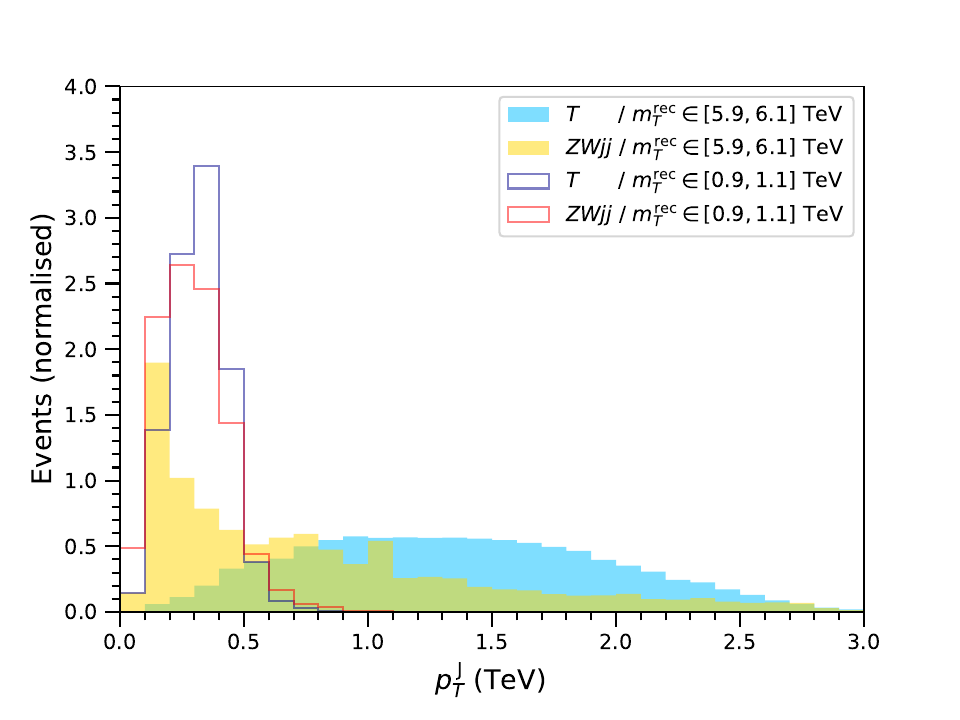} 
\end{tabular}
\caption{Kinematical distributions of $p_T^Z$ (top) and $p_T^J$ (bottom) for the signal and background, at different mass scales  $m_T^\mathrm{rec} \simeq 1$ TeV, 6 TeV.}
\label{fig:PT}
\end{center}
\end{figure}

Several kinematical distributions are quite different at low and high scales, not only for the signal but also for the background. We illustrate the differences in Fig.~\ref{fig:PT} with examples for $p_T^Z$ and $p_T^J$, at mass scales $m_T^\mathrm{rec} \simeq 1$ TeV, 6 TeV. For the lepton transverse momenta the behaviour is similar.
It is evident that the correlation between the actual background shape and the mass scale can aid in distinguishing signal from background. This is precisely the advantage of mass-unspecific classifiers, as we will see in the next section. Other distributions, however, do not change shape significantly from 1 TeV to 6 TeV, namely the reconstructed top quark mass and the forward jet multiplicity.

One can clearly see the reason why the $\mu$NN and $\mu$BDT improve over the pNN. When evaluated on signal and background events at $m_T \sim 6$ TeV, the mass label is set to $m_T \sim 6$ TeV. However, in the training some of the events to which the label $m_T \sim 6$ TeV was assigned, have features that do not correspond to that scale (see fig.~\ref{fig:PT} for two examples) and the discrimination is sub-optimal. Other features, however, are the same for all background scales, so the discrimination provided by the pNN is still reasonable.

\section{Mass-unspecific classifiers}

Following the MUST concept~\cite{Aguilar-Saavedra:2020uhm}, the event sample used to train the $\mu$NN and $\mu$BDT includes 5000 background events for each interval of $m_T^\mathrm{rec}$, from $[0.9,1.1]$ TeV up to $[6.3,6.5]$ TeV, totaling $1.4 \times 10^5$ events.\footnote{There is no limitation to the mass range in which the method is applied. Here we have chosen it to be sufficiently wide so as to illustrate the improvement brought by the method.} The same number of signal events is selected for each interval, generated with $T$ masses between 1 and 6.4 TeV. In order to prevent overtraining we use a validation sample built in a similar fashion, but with 3500 signal and 3500 background events per $m_T^\mathrm{rec}$ bin. The pNN also uses signal samples of 5000 and 3500 events for training and validation, respectively, for each $T$ mass.
 On the other hand, according to the pNN proposal~\cite{Baldi:2016fzo} the background samples are generated inclusively, with a posterior cut $m_T^\mathrm{rec} \geq 0.9$ TeV. For training and validation the number of background events is  $1.4 \times 10^5$  and $9.6 \times 10^4$, respectively. 

In order to test the influence of event sampling (balanced set across all the $m_T^\text{rec}$ range versus inclusive sampling, with or without event weights) we train a weighted BDT (wBDT) that uses $m_T^\mathrm{rec}$ as mass label as the $\mu$NN and $\mu$BDT do, but is trained on the inclusive background sample, using event weights. Provided training samples are very large, this should be equivalent to using a balanced set, i.e. with generation cuts. Conversely, we also train a pNN using a balanced set, which we dub as pNN$_B$.

We also train an extended pNN that, in addition to the mass label (true $T$ mass for the signal, random for the background), includes $m_T^\mathrm{rec}$ among the input features. This classifier is dubbed as pNN$_x$. The motivation for its inclusion is that some searches, for example Refs.~\cite{ATLAS:2023uox,ATLAS:2024uvu}, use classifiers designed in this fashion: not only they include the mass label, but also a feature that characterises the overall mass scale---and could thus be a replacement for the mass label. A summary of the similarities and differences between the classifiers used is given in Table~\ref{tab:summ}.
 
 \begin{table*}[htb]
 \begin{center}
 \begin{tabular}{ccccc}
 & training range & signal gen.  & background gen. & mass label \\
 $\mu$NN, $\mu$BDT & $[0.9,6.5]$ TeV & 0.2 TeV steps & 0.2 TeV bins & $m_T^\mathrm{rec}$ \\
 wBDT & $[0.9,6.5]$ TeV & 0.2 TeV steps & inclusive, weighted & $m_T^\mathrm{rec}$ \\
 pNN & $[0.9,6.5]$ TeV & 0.2 TeV steps & inclusive, unweighted & $m_T$ / random \\
 pNN$_B$ & $[0.9,6.5]$ TeV & 0.2 TeV steps & 0.2 TeV bins & $m_T$ / random \\
 pNN$_x$ & $[0.9,6.5]$ TeV & 0.2 TeV steps & inclusive, unweighted & ($m_T$ / rand.), $m_T^\mathrm{rec}$ \\
 $\mu\text{BDT}_\text{1000}$ & $[0.9,1.1]$ TeV & $m_T = 1$ TeV & $[0.9,1.1]$ TeV & $m_T^\mathrm{rec}$ \\
 $\mu\text{BDT}_\text{6000}$ & $[5.9,6.1]$ TeV & $m_T = 6$ TeV & $[5.9,6.1]$ TeV & $m_T^\mathrm{rec}$
 \end{tabular}
 \caption{Summary of similarities and differences between the classifiers used.}
 \label{tab:summ}
 \end{center}
 \end{table*}

For the training of the multivariate discriminators we use the kinematical variables listed in the previous section, properly standardised in order to improve the efficiency and convergence of the training.
The $\mu$NN, pNN, pNN$_x$ and pNN$_B$ are implemented using Keras \cite{chollet2015} with a TensorFlow backend \cite{tensorflow2015}.
They have two hidden layers of 64 nodes each with Rectified Linear Unit (ReLU) activation for the hidden layers and a sigmoid function for the output one. The NN optimisation relies on the binary cross-entropy loss function, using the Adam~\cite{Adam} algorithm. We perform 5 trainings with different initial seeds, and select the one that gives the highest area under the receiver operating characteristic curve (AUC) for the validation sample. Since the training samples are quite large, the spread in the AUC between trainings is at the $10^{-3}$ level.
The $\mu$BDT and wBDT are built using eXtreme Gradient Boosting ({\scshape XGBoost})~\cite{Chen:2016:XST:2939672.2939785}, with a maximum of 500 boosting trees and a depth of 5, a learning rate of 0.15, and otherwise default parameters. Here we also perform 5 trainings and select the classifier with the highest AUC.

The different discriminators are tested for  two $T$ masses $m_T = 1$ and 6 TeV, spanning a $m_T^\mathrm{rec}$ interval of 200 GeV around $m_T$. In addition to the classifiers mentioned before, we consider specific $\mu$BDTs trained only in the interval of $m_T^\mathrm{rec}$ tested (instead of the full interval $[0.9,6.5]$ TeV). The details of these classifiers are also included in Table~\ref{tab:summ}.

\begin{figure}[t]
\begin{center}
\begin{tabular}{c}
\includegraphics[height=6cm,clip=]{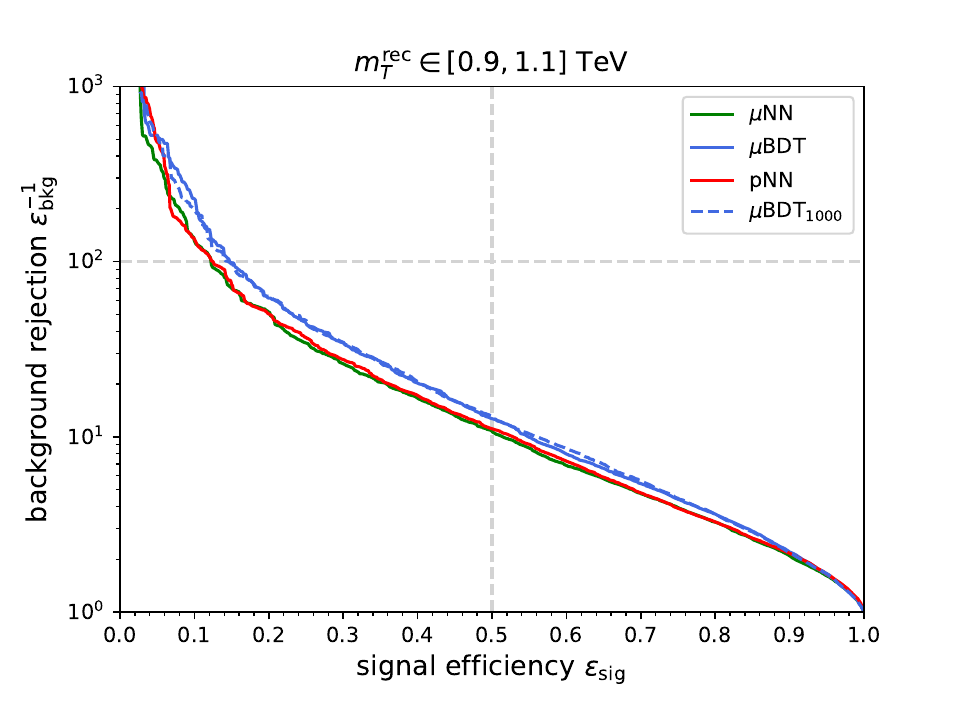} \\[1mm]
\includegraphics[height=6cm,clip=]{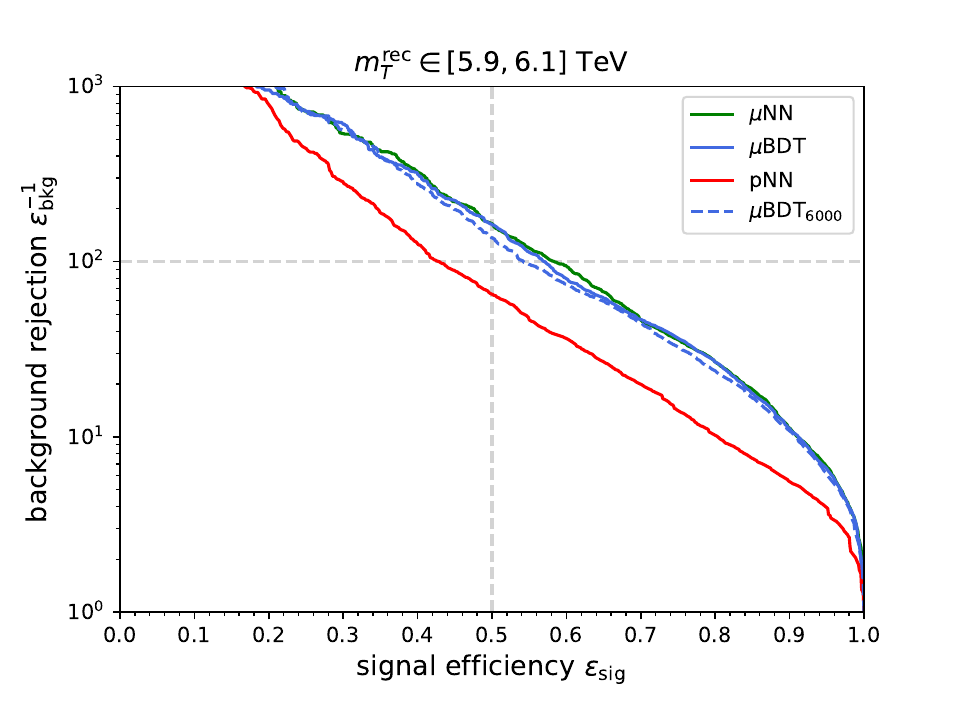} 
\end{tabular}
\caption{ROC curves for several classifiers, for $m_T = 1$ TeV (top) and $m_T = 6$ TeV (bottom).}
\label{fig:ROC}
\end{center}
\end{figure}

Figure~\ref{fig:ROC} shows the receiver operating characteristic (ROC) curves for the $\mu$NN, $\mu$BDT, pNN, and the specific classifiers, evaluated for $m_T = 1$ TeV (top) and $m_T = 6$ TeV (bottom). 
In all cases the mass-unspecific discriminators are equal or better than the pNN. 
For the lower mass the performance is close, with a small advantage for the BDTs. For the higher mass the $\mu$NN and $\mu$BDT largely improve over the pNN, both at low and high background rejections. For example, at $\varepsilon_\mathrm{bkg}^{-1} = 100$ the signal efficiency is a factor of 1.7 larger.
This is expected because the background $p_T$ distributions drastically differ at low and high masses (see Fig.~\ref{fig:PT}), and the pNN cannot capture their correlation across all the mass range. Nevertheless, the pNN still performs reasonably well, since some of the variables used (e.g. $m_t^\mathrm{rec}$ and the forward jet multiplicity) remain relatively stable across mass scales. 
Incidentally, the training times of the pNN and $\mu$NN are quite similar, respectively $345\,s$ and $330\,s$ in the same platform, while the BDT training only takes $12\,s$.

Notably, the performance of mass-unspecific discriminators, trained on the $[0.9,6.5]$ TeV range, is almost optimal in the sense that it equals the performance of mass-specific ones. For $m_T^\text{rec} \sim 1$ TeV, the dedicated discriminator $\mu\mathrm{BDT}_{1000}$ trained in the interval $m_T^\mathrm{rec} \in [0.9,1.1]$ TeV has the same performance as the $\mu$BDT trained in all the mass range. And the same can be said about $\mu\mathrm{BDT}_{6000}$ versus $\mu$BDT, when evaluated for $m_T^\text{rec} \sim 6$ TeV.

\begin{figure}[t]
\begin{center}
\begin{tabular}{c}
\includegraphics[height=6cm,clip=]{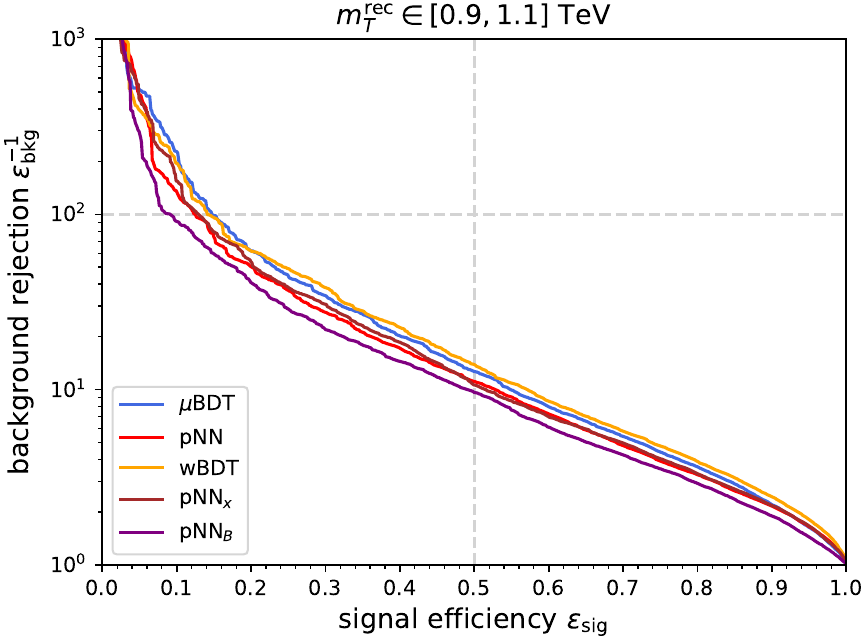} \\[1mm]
\includegraphics[height=6cm,clip=]{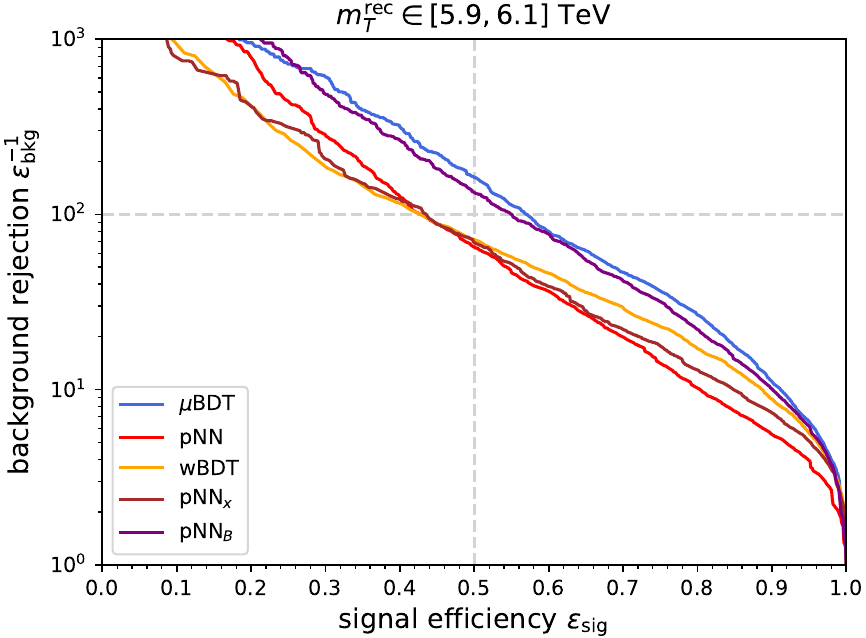} 
\end{tabular}
\caption{ROC curves for several classifiers, for $m_T = 1$ TeV (top) and $m_T = 6$ TeV (bottom).}
\label{fig:ROCb}
\end{center}
\end{figure}

In order to break down the different components that contribute to the superior performance of the mass-unspecific classifiers, Fig.~\ref{fig:ROCb} shows the ROC curves for the $\mu$BDT and pNN (as extreme cases in the previous plots), as well as the wBDT, pNN$_x$, and pNN$_B$. It is seen that
\begin{itemize}
\item The wBDT, which uses $m_T^\mathrm{rec}$ as mass label but the inclusive background sample with event weights, has
similar performance as the $\mu$BDT at $m_T = 1$ TeV, but worse at $m_T = 6$ TeV. This difference highlights the fact that, even with training datasets of this size ($2.8 \times 10^5$ events, out of several million generated events), balancing the sample at generation leads to improved performance compared to event weighting. 
\item The pNN$_B$, trained on a balanced set, has much better performance than the pNN at higher masses, but also worse at $m_T = 1$ TeV---where it is outperformed by the $\mu$BDT, e.g. with a signal efficiency 1.6 times larger for $\varepsilon_\mathrm{bkg}^{-1} = 100$. The difference between the pNN and pNN$_B$ also illustrates the advantage of a balanced set, while the difference between the pNN$_B$ and $\mu$BDT shows that, even for a balanced set, the use of $m_T^\text{rec}$ as mass label allows for a better correlation between the signal and background features across the full mass scale.
\item The pNN$_x$ has quite similar performance to the pNN for background rejections $\varepsilon_\mathrm{bkg}^{-1} \lesssim 100$. This is in agreement with previous findings.
\end{itemize}

Overall, it is seen that the mass-unspecific classifiers $\mu$NN and $\mu$BDT have the best performance across all the mass range, with some advantage for the $\mu$BDT. As aforementioned, the advantage stands from the use of a balanced set in the training, and $m_T^\text{rec}$ as mass label for both the signal and background.

\section{Application to new physics searches}

\begin{figure}[t]
\begin{center}
\includegraphics[height=6cm,clip=]{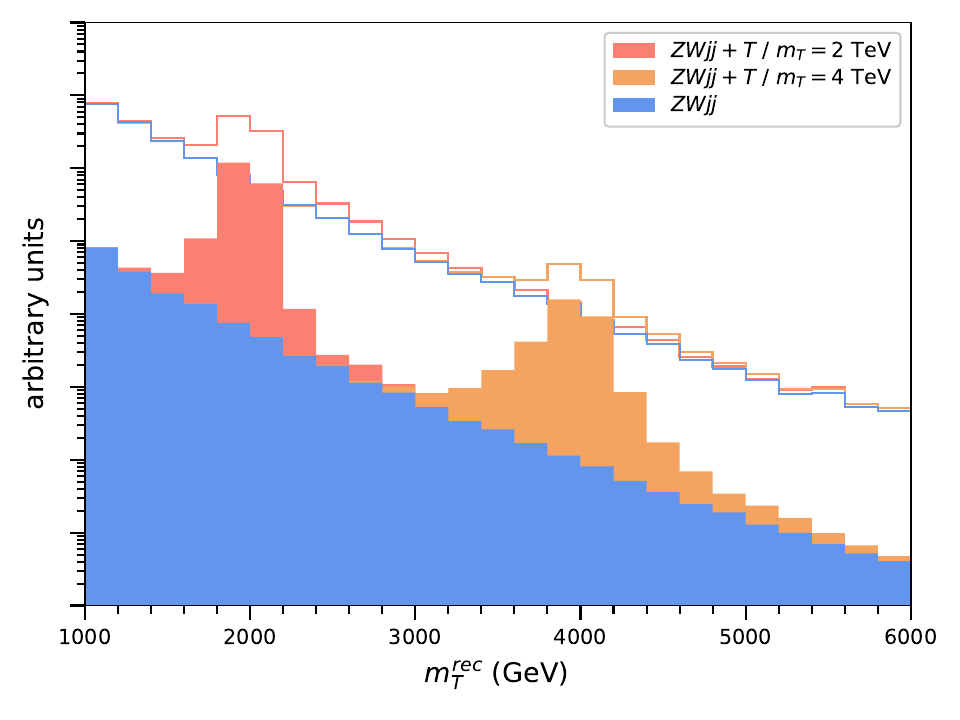} 
\caption{Kinematical distributions of the $ZW^+jj$ background, and with two injected $T$ signals with $m_T = 2$ and 4 TeV. The upper lines correspond to the preselection level, and the lower, filled distributions are obtained after a selection corresponding to $\varepsilon_\mathrm{bkg}^{-1} = 100$.}
\label{fig:mTrec}
\end{center}
\end{figure}

In analogy to mass decorrelation for jet taggers, it is possible to preserve the background shape after the application of mass-unspecific classifiers by the simple method of a varying threshold. An example is shown in Fig.~\ref{fig:mTrec}, for $T$ masses of 2 and 4 TeV, using the $\mu$BDT classifier. The same results are obtained with the $\mu$NN. Such event selection, with varying threshold but fixed background rejection as a function of $m_T^\mathrm{rec}$, enhances the visibility and significance of the signal peaks, whose shape is also preserved. Therefore, a precise background normalisation is not essential in order to observe the presence of a localised excess over a smooth background.
Figure~\ref{fig:mTrec} also shows that, despite being trained in bins of $m_T^\mathrm{rec}$, the $\mu$BDT produces a continuous output across bins.

\begin{figure}[t]
\begin{center}
\includegraphics[height=6cm,clip=]{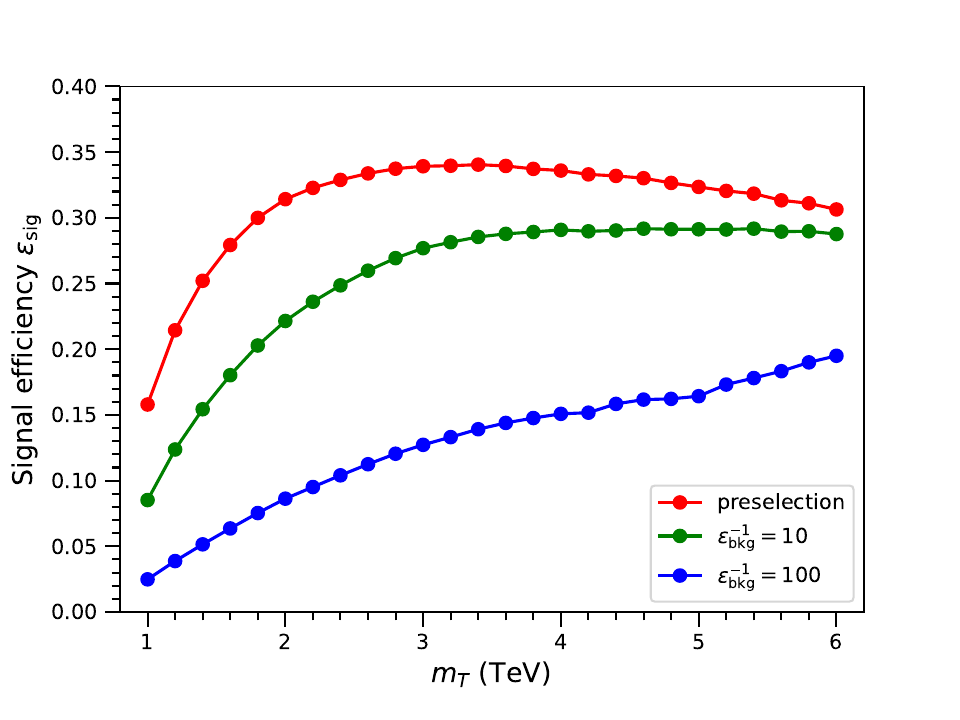} 
\caption{Signal efficiency at the preselection level and after selections on the $\mu$BDT score.}
\label{fig:effT}
\end{center}
\end{figure}

Fig.~\ref{fig:effT} presents the signal efficiency as a function of the $T$ quark mass, at the preselection level and after selections on the $\mu$BDT score corresponding to $\varepsilon_\mathrm{bkg}^{-1} = 10$ and 100. At the preselection level the efficiency is smaller at lower masses because often the top quark is not sufficiently boosted so that the $b$ quark and charged lepton are contained in a $R=0.8$ jet.\footnote{Experimental searches usually consider a `merged' as well as a `resolved' event selection to encompass the cases where the decay products are more or less boosted. This complication is out of the scope of the current work.}
At high masses the signal efficiency decreases due to the $b$ tagging. The selection on the $\mu$BDT score has an increasing efficiency for the signals because, as it can already be seen in Fig.~\ref{fig:ROC}, the signal to background discrimination improves at higher masses.

\section{Discussion}

In this paper we have built mass-unspecific classifiers  ($\mu$NN and $\mu$BDT)  that use the full correlation between the signal and background features and the mass scale explored. We have compared the $\mu$NN and $\mu$BDT, which produce a continuous output across all the mass range, to classifiers trained on a specific narrow mass interval. It is found that the performance of the $\mu$NN and $\mu$BDT is quite close to those specific classifiers. In this sense, the performance of mass-unspecific classifiers is nearly optimal, with a little advantage for the $\mu$BDT.

We have also seen that the $\mu$NN and $\mu$BDT perform as well as or better than a pNN and several variants of it. The advantage of mass-unspecific classifiers arises from two key factors in their design:
\begin{itemize}
\item The background mass scale is correlated with the actual background shape, enabling a more effective background identification at all mass scales.
\item The training sample includes events spanning the entire mass scale, setting equal weight for high as well as for low scales, thereby allowing the classifier to learn those differences.
\end{itemize}
We have compared the mass-unspecific classifiers to two variants of the pNN, either (1) adding $m_T^\text{rec}$ as input feature, or (2) using a balanced set for training. We have found that, overall, the performance of the $\mu$BDT and $\mu$NN is better. We have also tested a wBDT trained with inclusive sample and event weights, also finding a worse performance than the $\mu$BDT.

In our comparison of different classifiers we have considered a single background process. Based on prior experience with a multi-class jet tagger using the MUST method \cite{Aguilar-Saavedra:2021rjk}, we expect our conclusions to remain valid in the presence of multiple background processes. The underlying reason for the improvement achieved with mass-unspecific classifiers, namely the correlation between the background scale and its actual shape, continues to apply in such scenarios.

An additional point warrants discussion. We have tested the performance of the different classifiers on narrow intervals of the mass scale $m_T^\mathrm{rec}$. Tests using the full mass range $[0.9,6.5]$ TeV would be dominated by the behaviour at low masses, where most of the background events lie, and are insensitive to performance differences at high masses. This performance is precisely the one that determines how well a high-mass signal will be discriminated from high-mass background. Our approach also agrees with what most of the experimental searches using a pNN do: apply a selection on the discriminator and later use the mass variable to spot the presence of a signal.

Finally, we note that, despite the use of a benchmark of single $T$ production at the HL-LHC, our conclusions extend to other new physics processes and colliders. Mass-unspecific classifiers offer a good performance and are especially useful when a wide mass range can be explored.

\section*{Acknowledgements}

This work has been supported by the Spanish Research Agency (Agencia Estatal de Investigaci\'on) through projects PID2022-142545NB-C21, CEX2020-001007-S and grant PRE2022-101860 funded by MCIN/AEI/10.13039/501100011033.

\end{document}